\documentclass[twocolumn,english,showpacs,aps]{revtex4}
\usepackage[T1]{fontenc}
\usepackage[latin1]{inputenc}
\usepackage{graphicx}

\makeatletter

\usepackage{graphicx}

\usepackage{graphicx}

\usepackage{babel}
\makeatother
\begin{document}

\title{Calculating the nuclear mass at finite angular momenta}

\author{B.G.\ Carlsson and I.\ Ragnarsson}

\affiliation{Division of Mathematical Physics, Lund Institute of Technology, P.O.
Box 118, SE-221 00 Lund, Sweden}

\begin{abstract}
Mean field methods to calculate the nuclear mass are extended into
the high-spin regime to cal\-culate the nuclear binding energy as
a function of proton number, neutron number and angular momen\-tum.
Comparing the trend as a function of mass number for a selection of
high-spin states, a similar agreement between theory and experiment
is obtained as for ground state masses. 
\end{abstract}


\maketitle
A fundamental property of nuclei is their mass or equivalently, their
binding energies, $\mathcal{B}$. The variation of the nuclear mass
with proton and neutron number will reveal the shell effects which
are closely related to the magic numbers and the extra binding associated
with these numbers. It will also give some general idea about which
regions of nuclei are deformed and if some specific particle numbers
gives rise to extra binding for deformed nuclear shapes. In recent
years, it has become possible to study a large number of nuclei up
to very high angular momenta. A natural extension is then to study
the variation of the total nuclear energy as a function of the angular
momentum, $I$, i.e.\ to extend the investigations of the binding
energy to three dimensions, ${\mathcal{B}}(Z,N,I)$.

A first attempt to study the experimental shell effects at high spin
was carried out in Ref.\ \cite{Exp.Shell} where the energies of
high-spin states in $Z=50-82$ nuclei were plotted relative to a somewhat
schematic rotating liquid drop energy. In the present study we will
instead investigate how well the high-spin states in a few selected
nuclei are reproduced by state-of-the-art macroscopic-microscopic
calculations. A big simplification at high spin is that pairing correlations
are negligible which should make our calculations more reliable. We
will thus consider a limited number of nuclei whose level schemes
are known up into the unpaired regime and which have been successfully
interpreted in calculations.

In the present study, we will start from the finite range version
of the liquid drop model because for this model systematic fits to
the masses of all nuclei with $N,Z\geq8$ have been performed \cite{FRLDM}.
It turns out however that this model has some problems at low mass
numbers where the energy becomes unstable with respect to high multipole
deformations \cite{L-O}. Therefore, we will also consider a recent
version of a classical liquid drop formula, the LSD model \cite{LSD}
where an $A^{1/3}$ curvature term has been added to the classical
Myers-Swiatecki expressions \cite{Mye66}.

In the macroscopic-microscopic approach, the total nuclear energy
is obtained as \begin{equation}
E_{\mathrm{tot}}=\min_{def}\left[E_{\mathrm{l.d.}}(def)+\frac{\hbar^{2}I(I+1)}{2\mathcal{J}_{\mathrm{rig.}}(def)}+E_{\mathrm{shell}}(def,I)\right]\label{totener}\end{equation}
 in the high-spin limit when pairing is ignored. In the formula, it
has been specified which terms depend on deformation ($def$) and
angular momentum, $I$, respectively. The two first terms correspond
to the rotating liquid drop energy and the third term is the shell
energy.

In order to calculate the shell effects at high spin, we will rely
on cranked Nilsson-Strutinsky (CNS) formalism with the modified oscillator
potential because it is only in this model \cite{CNS,Afa99} that
systematic high-spin calculations have been carried out for nuclei
in essentially all mass regions. In order to keep the number of parameters
as small as possible, we have used the so-called $A=110$ parameters
\cite{Afa99} for all nuclei. These parameters have been optimized
for nuclei with $A=100-150$ but should be approximately applicable
for all mass numbers.

As one alternative for the static liquid drop energy, $E_{\mathrm{l.d.}}$
in Eq.\  (\ref{totener}), we consider the finite range liquid drop
model (FRLDM) \cite{FRLDM} which is more consistent \cite{FRD not good}
than the finite range droplet model (FRDM) in the description of fission
and thus probably more reliable for strongly deformed shapes. The
FRLDM parameters have been fitted to reproduce ground state masses
for 1654 nuclei and 28 fission-barrier heights. The resulting root
mean square error after experimental errors have been compensated
for is $\sigma_{\mathrm{th}}=0.779$ MeV \cite{FRLDM}. In the present
study, this model can not be applied directly because pairing is important
at the ground state while we are interested in unpaired high-spin
states. Indeed, in the fit, an average pairing energy is included
in the macroscopic liquid drop energy and must be removed. Furthermore,
a zero-point energy for vibrations in the elongation direction is
also included which appears inconsistent for high-spins where the
special symmetry favoring axial shapes is broken. One might consider
to include the full quadrupole zero-point energy term but this would
be very difficult in practice. Thus, a new fit to the same masses
has been performed where the full pairing interaction has been included
in the microscopic terms and where no zero-point energy is included.
This new fit gives a very similar mean square error, namely $\sigma_{\mathrm{th}}=0.778$
MeV, corresponding to a `standard' root mean square error of 0.783
MeV. The biggest difference between the two fits is the constant term
which increases from 2.6 to 7.2 MeV but there are also minor differences
in all other terms.

\begin{figure}
\includegraphics[%
  clip,
  width=0.85\columnwidth,
  keepaspectratio]{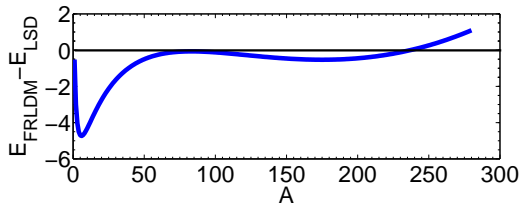}

\caption{\label{spherical diff}Difference in energy between the FRLDM and
the LSD model for nuclei along the beta-stability line at spin zero
and spherical shape.}
\end{figure}

The second alternative for the static liquid drop energy, the LSD
model \cite{LSD}, has been fitted to the same shell corrections as
the FRLDM for $Z\geq29$ and $N\geq29$ but for lighter nuclei semi-empirical
shell corrections are used instead. Therefore, as illustrated for
nuclei along $\beta$-stability in Fig.\ \ref{spherical diff},%
the two models give very similar energies for mass numbers $A\approx50-250$
but they become rather different for smaller mass numbers. The LSD
model has been fitted to 2766 masses with a root mean square error
of 0.698 MeV which decreases to 0.610 MeV when only 1654 masses are
considered. In order to remove the zero-point energy and the average
pairing when using the LSD model without refitting the parameters
we make an estimate based on the two fits for the FRLDM. We calculate
the energy difference for a spherical nuclei between the two fits
and add this contribution to the LSD model. This way we introduce
a constant shift for each nucleus but we do not change the original
deformation dependence in the model.

With the static liquid-drop energy fixed, it only remains to fix the
parameters of the rigid moment of inertia which enters in the second
term of Eq.\ (\ref{totener}). It can be calculated as \cite{Dav76}\begin{equation}
{\mathcal{J}_{\mathrm{rig.}}}=\frac{2}{5}Mr_{0}^{2}A^{2/3}\delta_{\mathcal{J}}(\varepsilon_{2},\gamma,\varepsilon_{4})+4Ma^{2}\label{inertia}\end{equation}
 where the radius for a spherical nucleus is parameterized as $r_{0}A^{1/3}$
and where the diffuseness is introduced in the form of a Yukawa folding
function with range $a$. A density distribution which changes from
10\% to 90\% of its central value in a distance of 2.4 fm is described
by $a\approx0.75$ fm. $M$ is the nuclear mass which varies as $A$.
The first expression on the right in Eq.\ (\ref{inertia}) is the
rigid body moment of inertia for a spherical nucleus with a sharp
boundary. The deformation correction, $\delta_{\mathcal{J}}(\varepsilon_{2},\gamma,\varepsilon_{4})$,
which equals one for spherical shape, is larger than one for shapes
mainly relevant for rapidly rotating nuclei, i.e.\ for rotation around
the smaller axis ($0^{\circ}<\gamma<60^{\circ}$). Note that the diffuseness
correction (the last term in Eq.\ \ref{inertia}) is independent
of deformation.

The moment of inertia parameters, $r_{0}$ and $a$ are obtained by
a fit to experimentally determined nuclear charge density distributions.
We fit the root-mean-square value of the radius $\langle r^{2}\rangle^{1/2}(\varepsilon_{2},r_{0},a)$,
calculated with a Yukawa folding function, to the values given in
Ref.\ \cite{e- scatter} for 116 nuclei with $A>16$. The quadrupole
deformations of the ground-states are taken from Ref.\  \cite{FRLDM}.
The result of the fit is $r_{0}=1.1599$ fm and $a=0.5984$ fm and
the standard deviation of the errors are $s=0.0454$ fm. Since the
root-mean-square radius as well as the moment of inertia involves
integrals over $\langle x^{2}\rangle$, $\langle y^{2}\rangle$ and
$\langle z^{2}\rangle$, this should give accurate values for the
moment of inertia even though there might be a larger uncertainty
in the two values, $r_{0}$ and $a$.

\begin{figure}
\includegraphics[%
  clip,
  width=1.0\columnwidth,
  keepaspectratio]{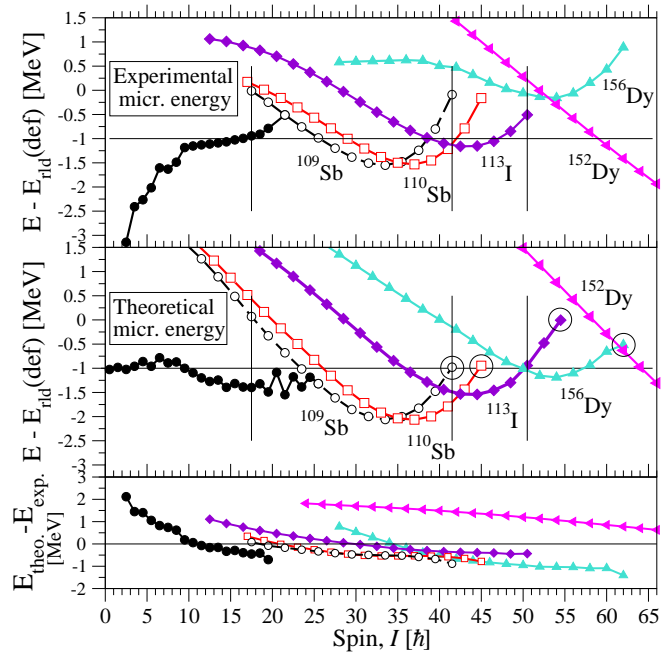}

\caption{\label{expe.shell.energy}Experimental (top panel) and theoretical
(middle panel) microscopic energies and their difference (lower panel)
for some well-established high-spin bands in $A=109-156$ with the
level scheme of $^{109}$Sb extended down to $I=0$. The LSD model
and a diffuse surface ($r_{0}$=1.16 fm, $a$=0.6 fm) has been used
when calculating the first and second terms, respectively, of Eq.\ (\ref{totener}).
Calculated states where all valence nucleons have their spin vectors
aligned with the axis of rotation (terminating states) are encircled.}
\end{figure}

Fig.\ \ref{expe.shell.energy} illustrates the total energy for a
few nuclei as a function of spin. It is drawn using the LSD model
for the static liquid drop energies and radius constants $r_{0}$=1.16
fm and $a$=0.6 fm obtained from the fit described above. For each
nucleus the energy of the corresponding rotating liquid drop has been
subtracted. This is thus a straightforward generalization of `standard'
mass plots, see e.g.\ Figs.\ 1 and 2 of Ref.\ \cite{FRLDM}, which
gives us a convenient scaling when comparing the microscopic energy
for different nuclei or when comparing theory and experiment.

The yrast line for $^{109}$Sb in Fig.\ \ref{expe.shell.energy}
is drawn starting from low spins where the discrepancies between between
calculations and experiment should be an approximate measure of the
pairing energy which is not included in the calculations.%
\begin{figure}
\includegraphics[%
  width=1.0\columnwidth]{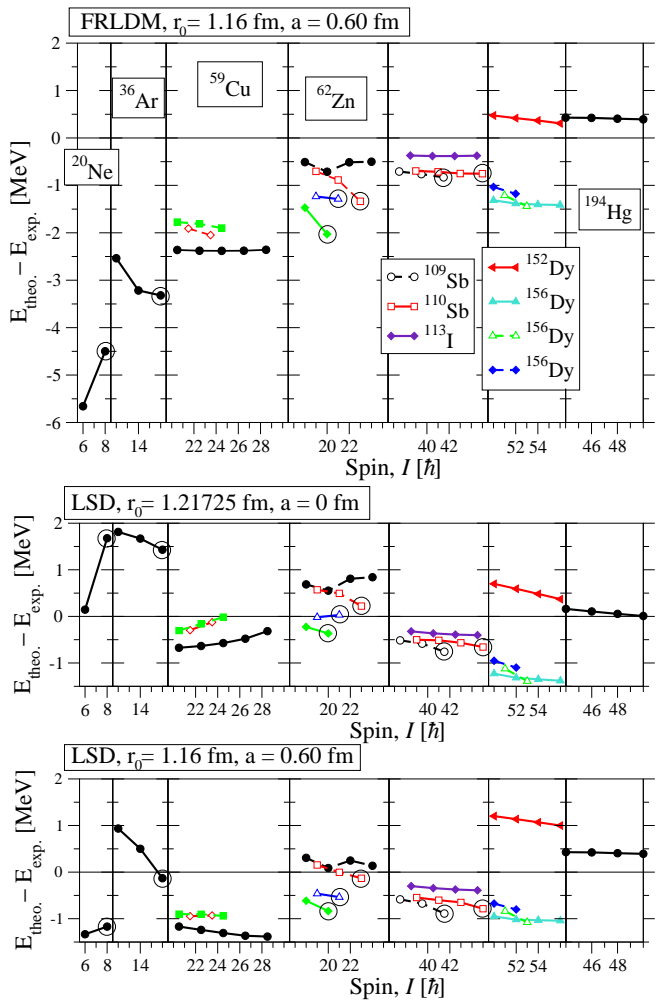}

\caption{\label{result}Difference in total energy between theory and experiment
for 10 nuclei with masses in the range $A=20-194$. In the lower panel,
the same parameters as in Fig.\ \ref{expe.shell.energy} have been
used while in the upper panel, the LSD model for the static liquid
drop energy has been replaced by the FRLDM and, in the middle panel
the diffuse surface when calculating the rigid moment of inertia has
been replaced by a sharp surface.}
\end{figure}
 The high-spin bands in $^{109,110}$Sb are both observed to termination
at $I=41.5$ and $I=45$, respectively. They are some of the first
and best developed smooth terminating bands in the $A=110$ region
\cite{Afa99}. Relative to the $^{100}$Sn core, they have two proton
holes in the g$_{9/2}$ shell and one proton and two and three neutrons,
respectively, in the high-$j$ h$_{11/2}$ orbitals. The band in $^{113}$I,
referred to as band 1 in Ref.\ \cite{Sta01}, has the same number
of holes and thus more valence particles which means that it has a
higher terminating spin value at $I^{\pi}=54.5^{+}$ and it is observed
up to $I=50.5$. The $^{156}$Dy band \cite{Kon98} shown in Fig.\ \ref{expe.shell.energy}
is yrast at the highest spin values. It is interpreted as having four
proton holes in the $Z=64$ core and is tentatively observed to its
termination at $I=62$. The band drawn for $^{152}$Dy is the first
superdeformed band identified \cite{Twi86} in the $A=150$ region
and the only such band whose absolute excitation energy is known \cite{Lau02}
. Its interpretation in CNS calculations is discussed in Ref.\ \cite{Rag93}
.

In order to extend the comparison of Fig.\ \ref{expe.shell.energy}
to a larger mass region, we compare calculated and experimental energies
for additional nuclei in Fig.\ \ref{result} where different versions
of the rotating liquid drop formula are used. Before discussing the
details of this figure, we will briefly describe the rotational bands
which in addition to the ones previously discussed have been included
in the comparison. The band observed to terminate at $8^{+}$ in $_{10}^{20}$Ne$_{10}$
\cite{Ale72} is built with two valence protons and two valence neutrons
in the d$_{5/2}$ shell. Cranked Nilsson-Strutinsky calculations for
this band have been discussed e.g.\ in Ref.\ \cite{Rag81} and reviewed
in \cite{Afa99}. The band in $^{36}$Ar is the so called superdeformed
band which is interpreted in CNS as well as shell model calculations
as having identical proton and neutron configurations with two particles
in f$_{7/2}$ and two holes in p$_{3/2}$ \cite{Sve00}. The highest
spin band shown for $^{59}$Cu \cite{And02} has been referred to
as superdeformed with four holes in the $^{56}$Ni core and consequently
seven particles in open shells with three of them in high-$j$ g$_{9/2}$
orbitals. The other band (with both signatures shown) has three holes
and two of the six particles in g$_{9/2}$ orbitals. The two bands
in $^{62}$Zn have one hole in the core and two or three of the seven
valence particles in the g$_{9/2}$ shell \cite{Sve98}. For $^{156}$Dy,
the two negative parity bands observed to highest spin, $I^{\pi}=52^{-},53^{-}$
are included in addition to the band shown in Fig.\ \ref{expe.shell.energy}.
Of the three superdeformed bands linked to the normal-deformed states
in the $A=190$ region, the band in $^{194}$Hg \cite{Kho96} has
been included in our comparison because it is observed to higher spins
than the other bands. According to the present calculations, it has
the high-$j$ $\pi(i_{13/2})^{4}\nu(j_{15/2})^{4}$ configuration
in agreement with standard interpretations, see e.g.\ \cite{Sat91}.

With the same parameters used in Fig.\ \ref{expe.shell.energy} and
in the lower panel of Fig.\ \ref{result}, those data points that
are repeated are identical. These parameters describe the data with
good accuracy. Typical errors are in the range of $\pm1$ MeV. Note
especially that from the limited data set considered here, the general
trend as a function of mass number appear correct. This is contrary
to the results in the upper panel where in the static liquid drop
energy, the LSD model has been replaced by the FRLDM. This results
in systematic differences for light nuclei where the calculated energies
are lower than the experimental energies. Thus, the FRLDM predicts
similar results as the LSD model for heavy nuclei but substantially
lower total energies for $^{20}$Ne, $^{36}$Ar and $^{59}$Cu. One
reason for this is that the two models are fitted to different shell
corrections for $Z<29$ and $N<29$ (as seen in Fig. \ref{spherical diff}).
Another reason is that for light nuclei the two models have different
deformation dependencies, with the FRLDM being softer in the $\varepsilon_{2}$
and $\varepsilon_{4}$ directions. Therefore for light nuclei the
FRLDM generally predicts somewhat larger $\varepsilon_{2}$ deformations
and in some cases so large $\varepsilon_{4}$ deformations that the
corresponding shapes appear unrealistic. Consider for example the
terminating state in $^{36}$Ar which is predicted to have a deformation
of $\varepsilon_{2}=0.415$ and $\varepsilon_{4}=0.188$ using the
FRLDM and $\varepsilon_{2}=0.356$ and $\varepsilon_{4}=0.065$ in
the LSD model. Similar results are obtained for the terminating state
in $^{20}$Ne which is predicted to have a deformation of $\varepsilon_{2}=0.120$
and $\varepsilon_{4}=0.210$ using the FRLDM and $\varepsilon_{2}=0.092$
and $\varepsilon_{4}=0.033$ in the LSD model.

The only difference between the two lower panels in Fig.\ \ref{result}
is that a sharp surface is used when calculating the rigid body moment
of inertia in the middle panel while a diffuse surface is used in
the lower panel. The radius in the sharp surface case is chosen as
the value used to calculate the Coulomb energy in the LSD formula,
$r_{0}=1.21725$ fm corresponding to a rigid moment of inertia of
a sphere which is about the same same as in the diffuse surface case
for $A\approx140$. In general, the differences between the two lower
panels are not so big for heavy nuclei but become more pronounced
for light nuclei. For example, the `diffuse moment of inertia' for
spherical shape is around 13 \% bigger for $A=36$ and more than 20
\% bigger for $A=20$. For $^{20}$Ne and $^{36}$Ar, this results
in an energy difference between 2.5 and 3 MeV for the terminating
$I=8$ and and $I=16$ states, respectively. For large deformations
however, these differences become smaller because the diffuseness
correction to the moment of inertia is independent of deformation
(Eq.\ \ref{inertia}). Thus, for the strongly deformed $16^{+}$
state in $^{36}$Ar, the difference is only about 1.5 MeV in the full
calculation while for the less deformed $8^{+}$ state of $^{20}$Ne
its around 3 MeV. Differences up to about 1 MeV are then seen in the
bands for the $A=60$ nuclei. An interesting effect is seen in the
superdeformed band in $^{152}$Dy. For spherical shape, the two formula
for the moment of inertia give energy differences (in opposite direction
relative to the light nuclei) of only around 0.2 MeV for $I=60$ but
the difference increases to 0.7 MeV at the calculated 2:1 deformation
of the superdeformed band.

With the present formalism we are thus able to put experimental and
theoretical results on a common absolute scale for high-spin calculations
with no pairing. In previous calculations, either the nuclear binding
energy at the ground state has been considered as a function of $N$
and $Z$ or the energy of a specific nucleus has been considered as
a function of angular momentum, $I$. Systematic trends or variations
of the microscopic energy at high spin might provide us with new insight
into high-spin phenomena, for example making it possible to compare
regions where angular momentum is built mainly from collective rotation
and mainly from single-particle excitations, respectively. A different
aspect is that the prediction of prompt particle decay from high-spin
states, which has been observed \cite{Rud02} for example in the nucleus
$^{59}$Cu discussed here, requires reliable estimates of the absolute
energy difference between the mother and daughter nucleus.

A natural extension of the present approach is to include many more
nuclei to see if the general trends of Fig.\ \ref{result} are still
the same. One could also make a new mass fit in the high-spin region
varying $r_{0}$ and $a$ in the moment of inertia formula or in the
long run, adjust `all parameters' in a global fit in the $N,Z,I$
space in which case it would clearly be necessary to also include
pairing. One possibility might be to calculate the pairing energy
only for $I=0$ and postulate a semiempirical formula how it slowly
disappears with angular momentum in which case it would be possible
to keep the configuration tracing in the high-spin regime which is
one of the most important features of the CNS approach. Especially
for light nuclei, the high-spin data appears important. For example,
it might be possible to obtain a more reliable estimate of the stiffness
towards deformation for nuclei with $A\lesssim100$ where fission
barrier data are unknown or uncertain but where the energy of several
strongly deformed high-spin bands are well established.

In summary, a consistent recipe has been introduced to consider the
total energy in the full $N,Z,I$ space. Good agreement between calculations
and experiment is obtained using a standard liquid drop expression
with an $A^{1/3}$ curvature term, an average moment of inertia calculated
from a diffuse surface mass distribution and shell corrections based
on the modified oscillator potential. For high-spin states in the
mass range $A=20-200$, the discrepancies appear comparable to those
obtained in state of the art mass calculations. The consequences of
using different models when calculating the rotating liquid drop energy
were investigated. 

The authors would like to thank Peter Möller for refitting the parameters
of the FRLDM. This work was supported by the Swedish Science Research
Council.

\end{document}